\documentclass[10pt,journal,compsoc,twoside]{IEEEtran}
\ifCLASSOPTIONcompsoc
  \usepackage[nocompress]{cite}
\else
  \usepackage{cite}
\fi
\ifCLASSINFOpdf
\else
\fi
\usepackage {graphicx}
\usepackage{subfigure}
\usepackage{latexsym}
\usepackage{setspace}
\usepackage{subfigure}
\usepackage{amsmath,amsfonts,amscd,amssymb}
\usepackage[noend]{alg-clipse}
\usepackage{alg-clipse}
\usepackage{multirow}
\usepackage{algorithm}
\usepackage{chapterbib}
\usepackage[none]{hyphenat}
\usepackage{url}
\usepackage{array}

\usepackage{chapterbib}
\usepackage{epstopdf}
\usepackage{caption}
\usepackage{setspace}
\usepackage{verbatim}
\usepackage{float}

\renewcommand{\algorithmicrequire}{\textbf{Input:}} 
\renewcommand{\algorithmicensure}{\textbf{Output:}} 
\usepackage{ragged2e}

\newtheorem{definition}{Definition}

\begin{document}
%
\title{Network Reconstruction and Controlling Based on Structural Regularity Analysis}
%
%
%
%

\author{Tao~Wu,~\IEEEmembership{}
        Shaojie~Qiao,~\IEEEmembership{}
        Xingping~Xian,~\IEEEmembership{}
        Xi-Zhao Wang,~\IEEEmembership{Fellow, IEEE, }
        Wei~Wang,~\IEEEmembership{}
        Yanbing~Liu

\IEEEcompsocitemizethanks{\IEEEcompsocthanksitem T. Wu is an Assistant Professor with School of Cybersecurity and Information Law, Chongqing University of Posts and Telecommunications, Chongqing 400065, China.

\IEEEcompsocthanksitem S. Qiao is a Professor with School of Cybersecurity, Chengdu University of Information Technology, Chengdu 610225, China. E-mail: sjqiao@cuit.edu.cn

\IEEEcompsocthanksitem X. Xian is a Ph.D. candidate with Department of Computer Science and Technology, Chongqing University of Posts and Telecommunications, Chongqing 400065, China. E-mail: xxp0213@gmail.com

\IEEEcompsocthanksitem X. Wang is a Professor with the College of Computer Science and Software Engineering, Shenzhen University, Shenzhen 518060, China.

\IEEEcompsocthanksitem W. Wang is an Associate Research Fellow with Institution of Cybersecurity, Sichuan University, Chengdu 610065, China.

\IEEEcompsocthanksitem Y. Liu is a Professor with Chongqing Engineering Laboratory of Internet and Information Security, Chongqing University of Posts and Telecommunications, Chongqing 400065, China.}

\thanks{Corresponding author: Shaojie~Qiao and Xingping~Xian.}}

\markboth{IEEE TRANSACTIONS ON KNOWLEDGE AND DATA ENGINEERING}
{WU \MakeLowercase{\textit{et al.}}: Network Reconstruction and Controlling Based on Structural Regularity Analysis}
%



\IEEEtitleabstractindextext{%
\begin{abstract}
\justifying
From the perspective of network analysis, the ubiquitous networks are comprised of regular and irregular components, which
makes uncovering the complexity of network structures to be a fundamental challenge. Exploring the regular information and identifying the roles of microscopic elements in network data can help us recognize the principle of network organization and contribute to network data utilization. However, the intrinsic structural properties of networks remain so far inadequately explored and theorised. With the realistic assumption that there are consistent features across the local structures of networks, we propose a low-rank pursuit based self-representation network model, in which the principle of network organization can be uncovered by a representation matrix. According to this model, original ``true" networks can be reconstructed based on the observed ``unreliable" network topology. In particular, the proposed model enables us to estimate the extent to which the networks are regulable, i.e., measuring the reconstructability of networks. In addition, the model is capable of measuring the importance of microscopic network elements, i.e., nodes and links, in terms of network regularity thereby allowing us to regulate the reconstructability of networks based on them. Extensive experiments on disparate real-world networks demonstrate the effectiveness of the proposed network reconstruction and regulation algorithm. Specifically, the network regularity metric can reflect the reconstructability of networks, and the reconstruction accuracy can be improved by removing irregular network links. Lastly, our approach provides an unique and novel insight into the organization exploring of complex networks.

\end{abstract}

\begin{IEEEkeywords}
Complex networks, organization principle, self-representation, network reconstruction, regularity controlling.
\end{IEEEkeywords}}

\maketitle

\IEEEdisplaynontitleabstractindextext

%
\IEEEpeerreviewmaketitle

\IEEEraisesectionheading{\section{Introduction}\label{sec:introduction}}

%
%
%
%
\IEEEPARstart{M}{any} real-world networked systems, including social system, biology system and Internet, are best thought of as a collection of discrete units that interact through a set of connections. Network paradigm~\cite{newman2003}, in which the discrete units or elements are denoted as network nodes and the connections are represented as link relationships, has been proved to be an ideal data form for exploring these systems recently. Driven by the increasing availability of data describing the structure of networks, the network data analysis community has seen a surge of interest in last twenty years, and the research focus has been transferred from statistical analysis based empirical studies~\cite{albert2002,barabasi2009,kossinets2006} to practical structure mining of networks, such as node ranking~\cite{lu2016vital,wu2017enhanced,wu2018power}, link prediction~\cite{liben2007link,lu2011link, wu2017predicting} and network reconstruction~\cite{Guimer2009Missing}. Exploring the essential components and uncovering the underlying principles of networks becomes critical for efficient network data utilization. Due to the inaccessibility of data sources, the subjectivity of participants as well as the limitation of data collection technologies, almost all of observed networks are unreliable with various noise level~\cite{Newman2017errors,Xiong2006Enhancing}. The noisy network data may cause fundamental wrong estimation or misleading conclusions in downstream analysis. Accordingly, network reconstruction and controlling that aims to infer the ``true'' underlying network and regulate the possibility of the inference becomes an elemental challenge. 

For network data utilization, network reconstruction and controlling via structural regularity analysis is of particular significance. Theoretically speaking, exploring the regularity of networks and identifying the roles of network elements in them can help to uncover their organization principles. Moreover, network regularity can be used to indicate the intrinsic predictability or reconstructability of networks. By measuring the regularity level of a network, we can determine whether the deficient performance of network reconstruction is caused by an inappropriate algorithm or is due to the irregularity of the network itself, and then estimate how large a space is there for performance improvement.
From the practical point of view, in online social networks, the potential commercial interests have led to the creation and proliferation of fake accounts, and it can help to find the fake accounts by detecting abnormal social relations via irregular links identification~\cite{jiang2014detecting}. In view of data mining applications, the presence of high noise level in networks can adversely impact the performance of data analysis. Hence, filtering out outlying links and estimating the ``true'' underlying network using reconstruction method is critical for data preprocessing~\cite{newman2018network,wang2018network}. In regard of data publishing, the sensitive relationships should be anonymized to protect private information, while link prediction methods are often used for sensitive links disclosure. Hence, regulating the predictability of networks based on critical links selection can be used for privacy-preserving~\cite{fard2012limiting}.

\subsection{\label{sec:level1}Motivations}

To solve the ``true'' underlying network inference problem, some works have been proposed in recent years, such as missing links prediction ~\cite{liben2007link,lu2011link}, network completion ~\cite{kim2011network}, network error estimation~\cite{newman2018network,newman2018pre} and network denoising~\cite{wang2018network}, all of which fall into the category of network reconstruction~\cite{Guimer2009Missing}. Most recently, the structural patterns mining problem in networks has received a lot of attention and many methods have been developed, including graphlet counting method~\cite{ahmed2015efficient,rossi2017estimation}, higher-order organizational pattern disclosure method~\cite{Benson2016Higher}, graph summarization method ~\cite{koutra2015summarizing}, and subgraph mining based graph classification method ~\cite{Wang2017Incremental,Vogelstein2013Graph}.
In fact, network structure can be partitioned into regular components and irregular components, in which only the former that
reflect the structural patterns of networks can be modeled and explained~\cite{L2015Toward}. Essentially, our capability of network
reconstruction depends on the regularity level of networks, i.e., the proportion of the regular components.
Thus, exploring the regular components that embody the structural pattern of networks is a promising way for reliable and sophisticated network reconstruction. To the best of our knowledge, the traditional network reconstruction works have the following disadvantages:


1) Most of the existing methods were proposed based on the prior assumptions on network organization, and they perform well only when the assumptions does hold. However, the real-world networks are often too complex to be depicted by using one specified mechanism. Till now, there is still little attention being paid on applying structural patterns learning in network reconstruction, which will cause a low reconstruction accuracy.



2) Due to the lake of technique of structural regularity analysis, traditional methods are not capable of measuring the intrinsic reconstructability of networks, which degrades the evaluation and the optimization on them.

3) Traditional algorithms focus on reconstructing networks as accurate as possible. Consequently, the intrinsic roles of substructures in macroscopic network organization are unclear, and eventually result in poor interpretability.

\subsection{\label{sec:level1}Our Work and Contributions}


In reality, complex networks are hard to be reconstructed if they tend to be random and changeable.
On the contrary, the networks can be reconstructed if they are of high regularity. In addition, network links play different roles in network organization and some of them have disproportionate influence on network regularity, and then network regularity can be regulated based on a limited number of critical network links.

Our research aims to uncover the ``true'' underlying network according to the unreliable observed network topology. We extend the traditional network reconstruction methods to measure the reconstructability of networks by exploring the organization principles of them, which can indicate the upper bound of reconstruction accuracy and provide guidance for algorithm optimization. By analyzing the roles of network elements, i.e., nodes and links, in network reconstruction, the reconstruction importance of them are defined and a network reconstructability controlling algorithm is proposed.


In summary, we make the following contributions:


\begin{itemize}
  \item Based on the assumption that networks contain the property of regularity and the data matrixes are approximately with low rank, we define a low rank pursuit based self-representation network model, in which networks can be represented as the linear combination of a few common structural bases and the high-order organization principle can be uncovered by the learned representation matrix. 

  \item By applying the self-representation network model in the observed networks, we propose a \textbf{L}ow \textbf{R}ank representation based \textbf{N}etwork \textbf{R}econstruction method (LRNR). By relaxing the low rank constraint of the model, we improve the LRNR method and propose a \textbf{L}ow \textbf{F}robenius norm based \textbf{N}etwork \textbf{R}econstruction algorithm (LFNR).

  \item According to the learned representation matrix, we define structural metrics to measure the regularity level of networks and the reconstruction importance of network elements. Based on them, the structure perturbation based network reconstructability regulation algorithm is proposed to filter out the irregular network links and promote network's potential
      for reconstruction.

\end{itemize}



The remainder of this paper is organized as follows. Section II surveys related work. Section III introduces the preliminaries
including importance definitions. Section IV provides the self-representation network model. Section V gives the algorithm for network reconstruction. Section VI presents our network regularity metric and reconstruction importance metric.
Section VII conducts experiments and Section VIII concludes the paper.


\section{Related Work}

\subsection{\label{sec:level1}Network modeling methods}


Most of the existing network inferring methods focus on discovering the missing network links~\cite{liben2007link, lu2011link}. The maximum likelihood estimation is a class of very commonly-used approaches in network modelling, including the stochastic block model (SBM)~\cite{Guimer2009Missing}, hierarchical structure model~\cite{Clauset2008Hierarchical}, etc., in which the model is selected based on the presupposed organizing rules of networks and the model parameters are learned from the observed structure of networks. Another kind of network inferring methods are the matrix based approaches. For example, Menon and Elkan integrated link prediction with collaborative filtering methods in recommendation and proposed matrix factorization based link prediction algorithm~\cite{Menon2011Link}. Wang et al.~\cite{Wang2016A} proposed a perturbation-based framework via negative matrix factorization (NMF) for missing links prediction.
Because of the noise and errors in networks, Zhou et al.~\cite{pech2017link} proposed a robust principal component analysis (RPCA) based link prediction algorithm which decomposes the adjacent matrix of networks into a low rank backbone structure and a sparse noise matrix. Qiao et al. \cite{qiao2018fast} analyzed the structures of networks and proposed a new approach based on the Mountain model for discovering large-scale communities. For the structure regulation of networks, Lu et al.~\cite{tan2016efficient} 
utilized the link prediction technique to generate exaggerated networks and identify critical nodes for network disintegration. Liu et al.~\cite{liu2015improving} proposed a metric of diffusion importance for network links and improved the method of computing the coreness centrality by filtering out redundant links.

In contrast to model based network reconstruction and structural pattern mining methods, we have combined them together to tackle the
complexity of networks for effective network reconstruction. Moreover, our work also focuses on the reconstructability of networks.

\subsection{\label{sec:level1}Low-rank Learning Method}

For processing big data in complex networks, a fundamental task is to find a low-dimensional representation of the high-dimensional data. To handle the problem,
principal component analysis (PCA)~\cite{Jolliffe1986Principal} was proposed to be one of most common approaches in recovering the best low-rank representation. However, the PCA method does work well for data with Gaussian noise, and its performance degrades for data with gross errors. Then, a more robust method robust principal component analysis (RPCA)~\cite{candes2011robust} was proposed, which can be formulated as follows.

\begin{spacing}{0.7}
\begin{equation}
\mathop {\min }\limits_{{\bf{A}},{\bf{E}}} rank({\bf{A}}) + ||{\bf{E}}|{|_1},\;s.t.,{\bf{X}} = {\bf{A}} + {\bf{E}}
\end{equation}
\end{spacing}

The PCA and RPCA methods assume that the data are distributed in one single space. Real-world data, however, often come from a set of multiple subspaces. To correctly partition the data into different subspaces, the Sparse Subspace Clustering (SSC)~\cite{Elhamifar2009Sparse} and Low Rank Representation (LRR)~\cite{Liu2010Robust} approaches were proposed. Formally, the SSC algorithm solves the following problem:

\begin{spacing}{0.7}
\begin{equation}\label{nt}
\mathop {\min }\limits_{\bf{Z},\bf{E}} ||\bf{Z}|{|_1} + {\lambda }||\bf{E}|{|_1}\;s.t.\,\bf{X} = \bf{XZ} + {\rm{\bf{E}}}\,{\rm{and}}\,{\rm{diag(\bf{Z}) = 0}}{\rm{.}}
\end{equation}
\end{spacing}

LRR is similar to SSC, except that it aims to find a low rank representation instead of a sparse representation. The objective function of LRR can be formulated as follows:
\begin{spacing}{0.7}
\begin{equation}\label{nt}
\mathop {\min }\limits_{\bf{Z},\bf{E}} ||\bf{Z}|{|_*} + \lambda ||\bf{E}|{|_{2,1}}\;s.t.\,\bf{X} = \bf{XZ} + {\rm{\bf{E}}}{\rm{.}}
\end{equation}
\end{spacing}

It is worthwhile to note that SSC figures out the sparse representation of each data vector individually, and it may not capture the global structure of $\bf{X}$. In contrast, LRR finds the lowest rank representation of all data jointly. 



\section{Preliminaries} \label{problem statement}

At its most fundamental level, the work contains the following essential phases: (1) model networks to explore their organization principle; (2) design algorithm to infer the ``true'' underlying networks; (3) estimate network regularity and the importance of network links; (4) perturb networks to change their regularity and reconstructability.

\begin{definition} [Network reconstruction] Given network $ G = (V, E) $, where $ V $ is the set of $|V|$ nodes and $ E \subseteq V \times V $ is the set of links, the goal of network reconstruction is to generate a reconstructed network based on an observed network ${G^T} = (V, {E^T}) $ to approximate the unknown underlying network $G$.
\end{definition}


For performance evaluation, the network $G$ is often used to construct ${G^T}$ via randomly adding and deleting links. All links of ${G^T}$ are denoted as training set ${E^T}$, ${E^T} = ({E} \backslash {E^M}) \cup {E^S} $, where ${E^M}$ and ${E^S}$ denote the missing links and the spurious links, respectively. The difference between the underlying network $G$ and the observed network ${G^T}$ is defined as testing set ${E^P}$, ${E^P} = {E^M} \cup {E^S} $.

\begin{definition} [Network regularity] The common structural characteristics across local structures of networks are defined as the organization principle of them. The network components that obey the principle are categorized to be regular, otherwise they are viewed to be irregular. The portion of the regular components of a network is defined as network regularity.
\end{definition}

\begin{definition} [Reconstruction importance] Changing the elements of networks, i.e., nodes and links, may alter the organization principle of the networks and thereby influence the accuracy of network reconstruction. Thus, the reconstruction importance of network elements can be estimated by analyzing their roles in the observed network structure.
\end{definition}

Due to the reasons such as privacy and confidentiality, we hope that the unknown network structures could not be inferred precisely based on the observations. Thus, it is necessary to develop methods to regulate the reconstructability of networks.

\begin{definition} [Network Reconstructability Regulation]
Given a network $G$, we perturb the network structure based on important link set $E^R$ to change its regularity level and obtain the final network denoted by ${{\rm{G}}^R}$, which has different reconstructability from the network $G$.
\end{definition}


An illustrative example of network reconstruction and controlling is shown in Fig. 1. The organization principle of the observed network is learned by the network representation model and then the reconstructed network is generated beyond it. The method can be evaluated by using the difference between reconstructed network and underlying network, i.e., residual links, as the testing set. As the level of regularity has a direct impact on the accuracy of reconstruction, the reconstructability of networks can be regulated by the method of irregular links based structural perturbation. The description of the notations used in this study are presented in Table 1.

\begin{table}
\begin{center}
\caption{Notations used in the paper.}
\begin{tabular}{|c|l|}
\hline
Notations & Descriptions \\
\hline
${\bf{X}}$ & The adjacency matrix of network  \\
\hline
${\bf{Z}}$ & The sparse matrix denoting the noise of ${\bf{X}}$ \\
\hline
$\lambda$ & The trade-off parameter \\
\hline
$G$ &  The ``true'' underlying network \\
\hline
$G^T$ &  The observed network corresponding to $G$  \\
\hline
$E^T$ &  The link set of $G^T$ used as training set \\
\hline
$E^P$ &  The difference between $G$ and $G^T$ \\
\hline
$E^M$ &  The missing link set  \\
\hline
$E^S$ &  The spurious link set  \\
\hline
$E^R$ &  The identified link set used for regulation \\
\hline
$G^R$ &  The regulated network generated from $G$ \\
\hline
$G^O$ &  The observed network corresponding to $G^R$ \\
\hline
$RC(k)$ &  The reconstruction importance of node $k$ \\
\hline
$U_{i,j}$ &  The reconstruction importance of link $(i,j)$ \\
\hline
${\sigma _r}$ &  The structure regularity of networks \\
\hline
\end{tabular}
\end{center}
\end{table}

\begin{figure*}[htb]
\centering
\includegraphics[width=7.0in, height =3.6in]{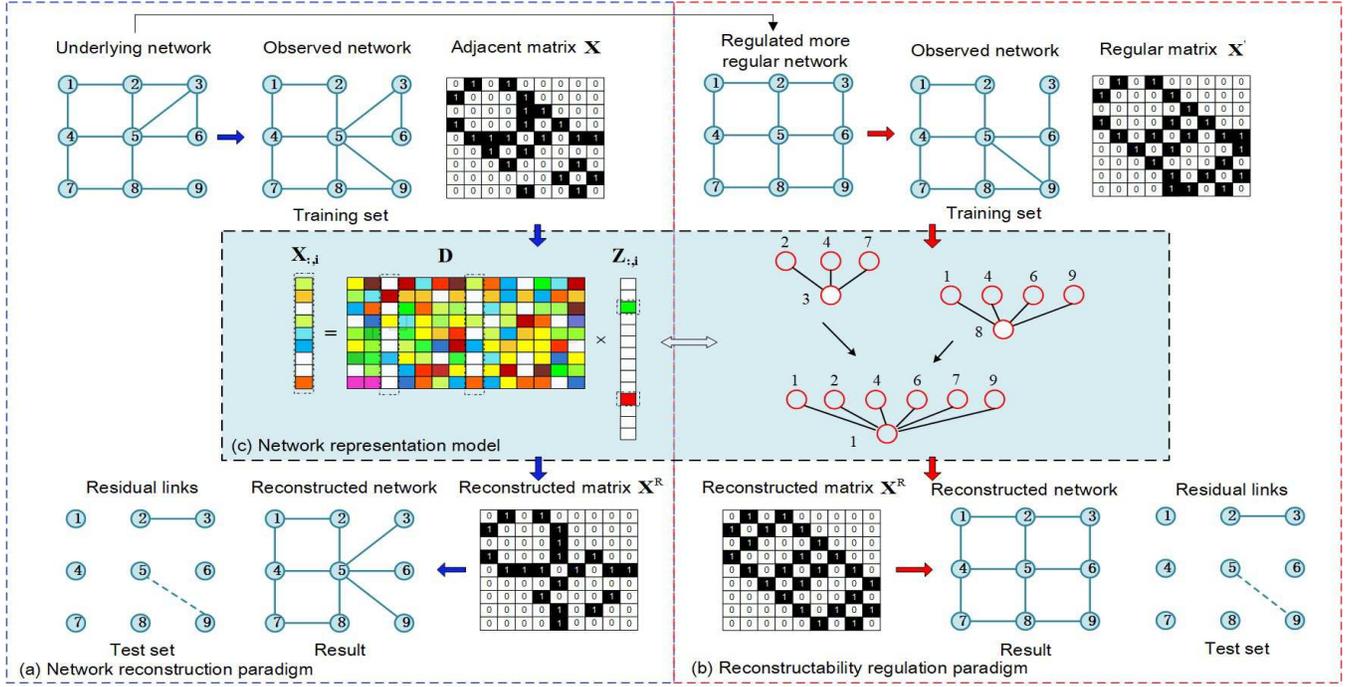}
\caption{Graphical illustration of network reconstruction and controlling. (a) Network reconstruction, in which the original network is reconstructed by applying the representation model in the observed network. (b) Reconstructability regulation, in which the intrinsic reconstruction accuracy of networks can be regulated by changing the level of regularity. (c) The low rank pursuit based network self-representation model.}
\label{fig_sim}
\end{figure*}

\section{\label{sec:level1}Self-representation network model}

Let $\bf{X} \in R^{n \times m}$ denote the adjacency matrix of the network $G$.
Each column of the matrix $\bf{X}$ is viewed as a local structure $\bf{X}_{:,i}$, thus $\bf{X}$ contains $m$ local structures, i.e.,
$[{\rm{\bf{X}}}_{:,1} ,{\rm{\bf{X}}}_{:,2} ,...,{\rm{\bf{X}}}_{:,m} ]$. Given a complete basis matrix ${\rm{\bf{D}}} = [{\rm{\bf{D}}}_{:,1} ,{\rm{\bf{D}}}_{:,2} ,...,{\rm{\bf{D}}}_{:,m} ] \in R^{n \times m}$, i.e., a collection of structural bases, each local structure ${\rm{\bf{X}}}_{:,i}$ can be represented by a linear combination of the bases, which is given as follows:
\begin{equation}\label{nt}
{{\bf{X}}_{:,i}} = {[{{\bf{D}}_{1,:}}{{\bf{Z}}_{:,i}},{{\bf{D}}_{2,:}}{{\bf{Z}}_{:,i}},...,{{\bf{D}}_{n,:}}{{\bf{Z}}_{:,i}}]^{\rm{T}}} = \sum\limits_{k = 1}^m {{{\bf{D}}_{:,k}}} {Z_{k,i}}
\end{equation}
\noindent where $Z_{k,i}$ corresponds to the weight of the base ${{{\bf{D}}_{:,k}}}$. That is, ${\rm{\bf{X}}}_{:,i}$ is actually the linear combination of matrix $\bf{D}$'s columns weighted by the elements of ${\rm{\bf{Z}}}_{:,i}$. Thus, the adjacency matrix $\bf{X} \in R^{n \times m}$ of network $G$ can be reconstructed by $\bf{X} = \bf{D}\bf{Z}$, where $\bf{Z} \in R^{m \times n}$ is a representation matrix.

To recognize the organization principle of networks, the best candidate for the basis matrix $\bf{D}$ is the adjacency matrix $\bf{X}$. Thus, each local structure ${\rm{\bf{X}}}_{:,i}$ can be represented as the combination of the others. Due to the fact that real-world networks have certain regularity and their local structures may have similar interaction patterns, the local structures can be reconstructed based on a common set of structural bases. That is, the columns of the representation matrix $\bf{Z}$ corresponding to similar local structures should be correlated, and $\bf{Z}$ is expected to be low-rank. In addition, the networks are always noisy and inaccurate. Based on the aforementioned discussion, networks can be modeled via low rank self-representation as follows:

\begin{spacing}{0.7}
\begin{equation}\label{nt}
\mathop {\min }\limits_{\bf{Z},\bf{E}} rank(\bf{Z}) + \lambda ||\bf{E}|{|_{2,1}}\;s.t.\,\bf{X} = \bf{XZ} + {\rm{\bf{E}}}{\rm{.}}
\end{equation}
\end{spacing}

\noindent where $\bf{E}$ is the error term, $\lambda \ge 0$ is a trade-off parameter to balance different terms. Here $l_{2,1}$ norm is adopted to characterize the node-specific corruptions.
Since this problem is NP hard, a common practice~\cite{Recht2007Guaranteed} is used to replace the rank of ${\bf{Z}}$ by its nuclear norm $||{\bf{Z}}|{|_*}$, i.e., the sum of its singular values, which leads to the following convex problem:

\begin{spacing}{0.7}
\begin{equation}\label{nt}
\mathop {\min }\limits_{\bf{Z},\bf{E}} ||\bf{Z}|{|_*} + \lambda ||\bf{E}|{|_{2,1}}\;s.t.\,\bf{X} = \bf{XZ} + {\rm{\bf{E}}}{\rm{.}}
\end{equation}
\end{spacing}

In order to solve the optimization problem (in Formula 6), we introduce auxiliary variable ${\bf{J}}$ to make the objective function separable. The problem can be converted to be:

\begin{spacing}{0.7}
\begin{equation}\label{nt}
\mathop {\min }\limits_{{\bf{Z}},{\bf{E}}} ||{\bf{J}}|{|_*} + \lambda ||{\bf{E}}|{|_{2,1}},\;s.t.,{\bf{X}} = {\bf{XZ}} + {\bf{E}},{\bf{Z}} = {\bf{J}}
\end{equation}
\end{spacing}

\noindent which can be handled by solving the following Augmented Lagrange Multiplier(ALM) problem~\cite{lin2010augmented}:

\begin{spacing}{0.6}
\begin{equation}\label{}
\begin{array}{l}
L({\rm{\bf{J}}}, {\rm{\bf{Z}}},{\rm{\bf{E}}}) = ||{\bf{J}}|{|_*} + \lambda ||{\bf{E}}|{|_{2,1}} + {\mathop{\rm tr}\nolimits} [{\bf{Y}}_1^{\mathop{\rm T}\nolimits} ({\bf{X}} - {\bf{XZ}} - {\bf{E}})] + \\  [2mm]
{\mathop{\rm tr}\nolimits} [{\bf{Y}}_2^{\mathop{\rm T}\nolimits} ({\bf{Z}} - {\bf{J}})] + \frac{\mu }{2}(||{\bf{X}} - {\bf{XZ}} - {\bf{E}}||_{\cal F}^2 + ||{\bf{Z}} - {\bf{J}}||_F^2)
\end{array}
\end{equation}
\end{spacing}

\noindent where ${\mathop{\rm \bf{Y}}\nolimits} _1$ and ${\mathop{\rm \bf{Y}}\nolimits} _2$ are Lagrange multipliers and $\mu  > 0$ is a penalty parameter. This problem can be solved by minimizing ${\rm{\bf{J}}}$, ${\rm{\bf{Z}}}$ and ${\rm{\bf{E}}}$, respectively. By taking the efficiency into consideration, we choose the inexact ALM method, which is shown in Algorithm 1. Notice that Step 3 can be solved via the singular value thresholding operator~\cite{Cai2008A}, and Step 5 can be solved according to the work~\cite{Liu2010Robust}.

\begin{algorithm}[!htb]
\caption{Solving Formula (8) by Inexact ALM method}
\begin{algorithmic}[1] \label{LRR}
  \INPUT adjacency matrix of observed network $\bf{X}$, trade-off parameter $\lambda$.
  \OUTPUT representation matrix ${\bf{Z}}$, error matrix ${\bf{E}}$.
  \STATE Initial $\bf{Z}=\bf{J}=\bf{E}=0$,$\bf{Y}_1=\bf{Y}_2=0$,$\mu={10^{ - 6}}$,${\max _\mu }={10^{10}}$, $\rho$  = 1.1, $\varepsilon  = {10^{ - 8}}$;

  \WHILE {not converged}
  \STATE Fix the others and update $\bf{J}$ by\quad \quad \quad \quad \quad \quad \quad \quad

  $\bf{J} = \arg \min ||\bf{J}|{|_*} + \frac{\mu }{2}||\bf{J} - (\bf{Z} + \frac{{{\bf{Y}_2}}}{\mu })||_{\cal F}^2$;

  \STATE Fix the others and update $\bf{Z}$ by\quad \quad \quad \quad \quad \quad \quad \quad

  $\bf{Z} = {({\bf{X}^{\rm T}}\bf{X} + \bf{I})^{ - 1}}({\bf{X}^{\rm T}}(\bf{X} - \bf{E}) + \;\bf{J} + ({\bf{X}^{\rm T}}{\bf{Y}_1} - {\bf{Y}_2})/\mu )$;

  \STATE Fix the others and update $E$ by\quad \quad \quad \quad \quad \quad \quad \quad

  $\bf{E} = \arg \min \lambda ||\bf{E}|{|_{2,1}} + \frac{\mu }{2} ||{\rm{\bf{E}}} - (\bf{X} - \bf{XZ} + \frac{{{\bf{Y}_1}}}{\mu }))||_{\cal F}^2$;

  \STATE Update the multipliers \quad \quad \quad \quad \quad \quad \quad \quad

  ${\bf{Y}_1} = {\bf{Y}_1} + \mu (\bf{X} - \bf{XZ}{\rm{ - \bf{E}}})$

  ${\bf{Y}_2} = {\bf{Y}_2} + \mu (\bf{Z} - \bf{J})$;

  \STATE Update the parameter $\mu$ by $\mu  = \min (\rho \mu ,{\max _\mu })$;

  \STATE Check the convergence conditions \quad \quad \quad \quad \quad \quad \quad \quad

  $||\bf{X} - \bf{XZ} - \bf{E}|{|_\infty } < \varepsilon$ and $|\bf{Z} - \bf{J})|{|_\infty } < \varepsilon$;

  \ENDWHILE

  \STATE \textbf{output} ${\bf{Z}}$, ${\bf{E}}$
\end{algorithmic}
\end{algorithm}


\begin{figure}[!t]
\centering
\includegraphics[width=3.5in,height =3.4in]{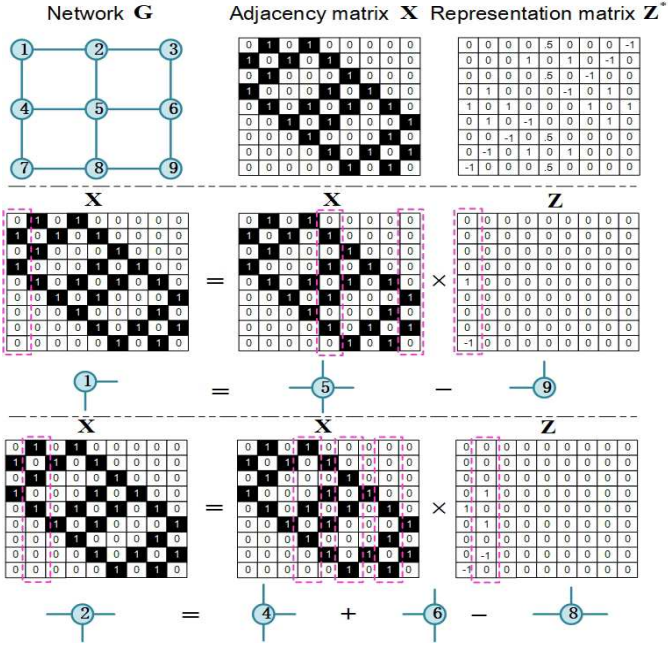}
\caption{Illustrating the self-representation network model. }
\label{fig_sim}
\end{figure}

\begin{figure}[!t]
\centering
\includegraphics[width=3.5in,height =0.99in]{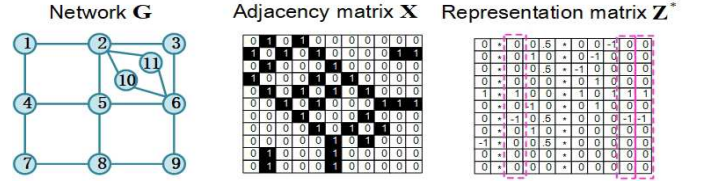}
\caption{An illustration of low-rank pursuit of representation matrix. The unknown entries are denoted by starts.}
\end{figure}

To explain the proposed self-represent network model, an example is given in Fig. 2, where the error term is specified to None. According to Fig. 2, the local structure of node $1$ is represented by that of node $5$ and node $9$. That is, the vector of matrix ${{\bf{X}}}$ corresponding to node $1$ is a linear combination of the columns of ${{\bf{X}}}$ weighted by the entries of the first column of the representation matrix ${{\bf{Z}}}$, where only the entries corresponding to node $5$ and node $9$ are nonzero. Similarly, the local structure of node $2$ can be represented by that of node $4$, $6$ and $8$. The final representation matrix ${{\bf{Z}}^{\rm{*}}}$ of this network is presented in the right of the first row. 

Furthermore, an example is given in Fig. 3 to illustrate the low rank pursuit of representation matrix ${{\bf{Z}}^{\rm{}}}$. By compared with Fig. 2, the common neighbors of node $2$ and $6$ include two new nodes $10$ and $11$, and the local structures of nodes $3$, $5$, $10$ and $11$ have high intercommunity. Thus, by figuring out the common interaction patterns among the nodes, one node's corrupted local structure can be reconstructed based on the structural atoms of the others. We can see that the network has higher structural redundancy for reconstruction than the network in Fig. 2. So, the low-rank pursuit of representation matrix ${{\bf{Z}}}$ (that requires its columns have similar values) actually means finding the principal patterns underlying the network.

\section{\label{sec:level1}Self-representation Model Based Network Reconstruction}

The goal of network reconstruction is to infer the ``true'' underlying network via finding the structural patterns of the observed network.
By applying self-representation network model $\mathop {\min }\limits_{{{\bf{Z}}},{{\bf{E}}}}||{{\bf{Z}}}|{|_*} + \lambda ||{{\bf{E}}}|{|_{2,1}},\;s.t.,{{\bf{X}}} = {{\bf{X}}}{{\bf{Z}}} + {{\bf{E}}}$ into the observed network ${G^T}$, the learned representation matrix ${{{\bf{Z}}^*}}$ reveals the organization principle of the network,
and the unknown structure can be inferred based on it. It is worth noting that
the core of the model is to learn the low rank representation of all units jointly to uncover the structural pattern of networks. However, the foremost concern of network reconstruction should be inferring underlying network in an accurate manner. Therefore, here we propose to replace the nuclear norm by the Frobenius norm and define the objective function as follows.
\begin{spacing}{0.7}
\begin{equation}\label{}
\begin{array}{l}
\mathop {\min }\limits_{{\bf{Z}},{\bf{E}}} ||{\bf{Z}}|{|_{\cal F}^2} + \lambda ||{\bf{E}}|{|_{2,1}},\;s.t.,{\bf{X}} = {\bf{XZ}} + {\bf{E}}\;
\end{array}
\end{equation}
\end{spacing}

By introducing auxiliary variables ${\bf{J}}$, the augmented Lagrangian function of the problem is converted as follows:

\begin{spacing}{0.7}
\begin{equation}\label{}
\begin{array}{l}
L({\rm{\bf{J}}}, {\rm{\bf{Z}}},{\rm{\bf{E}}}) = ||{\bf{J}}|{|_{\cal F}^2} + \lambda ||{\bf{E}}|{|_{2,1}} + {\mathop{\rm tr}\nolimits} [{\bf{Y}}_1^{\mathop{\rm T}\nolimits} ({\bf{X}} - {\bf{XZ}} - {\bf{E}})] \\ [3mm]
\quad  + {\mathop{\rm tr}\nolimits} [{\bf{Y}}_2^{\mathop{\rm T}\nolimits} ({\bf{Z}} - {\bf{J}})] + \frac{\mu }{2}(||{\bf{X}} - {\bf{XZ}} - {\bf{E}}||_{\cal F}^2 + ||{\bf{Z}} - {\bf{J}}||_{\cal F}^2)
\end{array}
\end{equation}
\end{spacing}

$\\$
\noindent \textbf{Lemma 1.}  \emph{For the matrix $\rm{\bf{X}}$, $\rm{\bf{Z}}$ and $\rm{\bf{E}}$, the augmented item ${\rm{tr}}[{\bf{Y}}_1^{\rm{T}}{\bf{(X - XZ - E)}}]$ and error item $||{\bf{X - XZ - E}}||_{\cal F}^2$ can be merged, i.e., we have the following equation:}

\begin{equation}\label{}
\begin{array}{l}
{\rm{arg}}\mathop {\min }\limits_{\bf{E}} {\rm{tr}}[{\bf{Y}}_1^{\rm{T}}{\bf{(X - XZ - E)}}]{\rm{ + }}\frac{\mu }{2}||{\bf{X - XZ - E}}||_{\cal F}^2    \\  [3mm]
 = {\rm{arg}}\mathop {\min }\limits_{\bf{E}} \frac{\mu }{2}||{\bf{X - XZ - E}} + \frac{{{{\bf{Y}}_1}}}{\mu }||_{\cal F}^2
\end{array}
\end{equation}


\noindent \textbf{Proof.} According to the definition of the interior product and the Frobenius norm of matrix, we have:
\begin{spacing}{0.7}
\begin{equation}\label{}
\begin{array}{l}
{\rm{arg}}\mathop {\min }\limits_{\bf{E}} {\rm{tr}}[{\bf{Y}}_1^{\rm{T}}{\bf{(X - XZ - E)}}]{\rm{ + }}\frac{\mu }{2}||{\bf{X - XZ - E}}||_{\cal F}^2\\  [3mm]
 = {\rm{arg}}\mathop {\min }\limits_{\bf{E}} {\rm{ < }}\frac{2}{\mu }{{\bf{Y}}_1},{\bf{X - XZ - E}}{\rm{ >  + }}||{\bf{X - XZ - E}}||_{\cal F}^2\\  [3mm]
 = {\rm{arg}}\mathop {\min }\limits_{\bf{E}}  < \frac{{{{\bf{Y}}_1}}}{\mu },\frac{{{{\bf{Y}}_1}}}{\mu } >  + 2{\rm{ < }}\frac{{{{\bf{Y}}_1}}}{\mu },{\bf{X - XZ - E}}{\rm{ >  + }}\\  [3mm]
\quad \quad   < {\bf{X - XZ - E}},{\bf{X - XZ - E}} >  -  < \frac{{{{\bf{Y}}_1}}}{\mu },\frac{{{{\bf{Y}}_1}}}{\mu } > \\  [3mm]
 = {\rm{arg}}\mathop {\min }\limits_{\bf{E}} ||{\bf{X - XZ - E}} + \frac{{{{\bf{Y}}_1}}}{\mu }||_{\cal F}^2 - ||\frac{{{{\bf{Y}}_1}}}{\mu }||_{\cal F}^2
\end{array}
\end{equation}
\end{spacing}

By removing the terms irrelevant to $\rm{\bf{E}}$, Formula 12 is converted to be:

\begin{spacing}{0.7}
\begin{equation}\label{}
 {\rm{arg}}\mathop {\min }\limits_{\bf{E}} ||{\bf{X - XZ - E}} + \frac{{{{\bf{Y}}_1}}}{\mu }||_{\cal F}^2
\end{equation}
\end{spacing}

Consequently, the equality does hold.

\begin{algorithm}[!htb]
\caption{Solving Formula (10) by Inexact ALM method}
\label{alg::conjugateGradient}
\begin{algorithmic}[1]
\INPUT adjacency matrix of observed network $\bf{X}$, trade-off parameter $\lambda$.
\OUTPUT representation matrix ${\bf{Z}}$, error matrix ${\bf{E}}$.
\STATE Initial $\bf{Z}=\bf{J}=\bf{E}=0$,$\bf{Y}_1=\bf{Y}_2=0$,$\mu={10^{ - 6}}$,${\max _\mu }={10^{10}}$, $\rho$  = 1.1, $\varepsilon  = {10^{ - 8}}$;

\WHILE {not converged}
\STATE Fix the others and update $\bf{J}$ by\quad \quad \quad \quad \quad \quad \quad \quad

$\bf{J} = \arg \min ||\bf{J}|{|_{\cal F}^2} + \frac{\mu }{2}||\bf{J} - (\bf{Z} + \frac{{{\bf{Y}_2}}}{\mu })||_{\cal F}^2$;

\STATE Fix the others and update $\bf{Z}$ by\quad \quad \quad \quad \quad \quad \quad \quad

$\bf{Z} = {({\bf{X}^{\rm T}}\bf{X} + \bf{I})^{ - 1}}({\bf{X}^{\rm T}}(\bf{X} - \bf{E}) + \;\bf{J} + ({\bf{X}^{\rm T}}{\bf{Y}_1} - {\bf{Y}_2})/\mu )$;

\STATE Fix the others and update $E$ by\quad \quad \quad \quad \quad \quad \quad \quad

$\bf{E} = \arg \min \lambda ||\bf{E}|{|_{2,1}} + \frac{\mu }{2} ||{\rm{\bf{E}}} - (\bf{X} - \bf{XZ} + \frac{{{\bf{Y}_1}}}{\mu }))||_{\cal F}^2$;

\STATE Update the multipliers \quad \quad \quad \quad \quad \quad \quad \quad

${\bf{Y}_1} = {\bf{Y}_1} + \mu (\bf{X} - \bf{XZ}{\rm{ - \bf{E}}})$

${\bf{Y}_2} = {\bf{Y}_2} + \mu (\bf{Z} - \bf{J})$;

\STATE Update the parameter $\mu$ by $\mu  = \min (\rho \mu ,{\max _\mu })$;

\STATE Check the convergence conditions \quad \quad \quad \quad \quad \quad \quad \quad

$||\bf{X} - \bf{XZ} - \bf{E}|{|_\infty } < \varepsilon$ and $|\bf{Z} - \bf{J})|{|_\infty } < \varepsilon$;

\ENDWHILE

\end{algorithmic}
\end{algorithm}

To solve Formula 10, we update each variable while fixing the others. To update variable ${\rm{\bf{J}}}$, by ignoring the irrelevant terms w.r.t. ${\rm{\bf{J}}}$ in Formula 10, we have the objective as follows:

\begin{spacing}{0.7}
\begin{equation}\label{}
\begin{array}{l}
{\bf{J}} = \arg \,\mathop {\min }\limits_{\bf{J}} \,||{\bf{J}}||_{\cal F}^2 + {\rm{tr}}[{\bf{Y}}_2^{\rm{T}}{\bf{(Z - J)}}]{\rm{ + }}\frac{\mu }{2}||{\bf{Z - J}}||_{\cal F}^2
\end{array}
\end{equation}
\end{spacing}

According to Lemma 1, we can combine ${\rm{tr}}[{\bf{Y}}_2^{\rm{T}}{\bf{(Z - J)}}]$ and $\frac{\mu }{2}||{\bf{Z - J}}||_{\cal F}^2$ and Formula 14 can be converted into:

\begin{spacing}{0.7}
\begin{equation}\label{}
\begin{array}{l}
 {\bf{J}} = \arg \,\mathop {\min }\limits_{\bf{J}} \,||{\bf{J}}||_{\cal F}^2 + \frac{\mu }{2}||{\bf{Z}}{\rm{ - }}{\bf{J}} + \frac{{{{\bf{Y}}_2}}}{\mu }||_{\cal F}^2\\   [3mm]
 = \arg \,\mathop {\min }\limits_{\bf{J}} \,{{\bf{J}}^{\rm{T}}}{\bf{J}} + \frac{\mu }{2} < {\bf{J}}{\rm{ - }}({\bf{Z}} + \frac{{{{\bf{Y}}_2}}}{\mu }),{\bf{J}}{\rm{ - }}({\bf{Z}} + \frac{{{{\bf{Y}}_2}}}{\mu }) >
\end{array}
\end{equation}
\end{spacing}

By specifying the derivative w.r.t. ${\bf{J}}$ to zero, we obtain:

\begin{equation}\label{}
{\bf{J}} = \frac{\mu }{{\mu  + 2}}({\bf{Z}} + \frac{{{{\bf{Y}}_2}}}{\mu })
\end{equation}

By applying Lemma 1, we can perform the update operation of ${\rm{\bf{E}}}$ that is similar to Algorithm 1. Meanwhile, the update operation of ${\rm{\bf{Z}}}$ remain unchanged. And, the problem in Formula 10 is solved by Algorithm 2.

\textbf{Network reconstruction}. By learning the optimal representation matrix ${\bf{Z}^{*}}$ of observed networks, the existence
likelihoods of network links can be inferred by combining matrix ${{{\bf{Z}}^*}}$ with the basis matrix ${\bf{X}}$, i.e.,



\begin{equation}\label{}
\begin{array}{l}
\bf{SM} = {\bf{X}}{\bf{Z}^*} + {{({\bf{X}}{\bf{Z}^*})}^{\rm T}}.
\end{array}
\end{equation}

Actually, the feasibility of the proposed network reconstruction method is based on the consistent patterns across local structures, in which the corrupted local structure can be rectified based on the feature of similar ones. All non-observed links are ranked according to their likelihoods, in which the links with high scores have a higher possibility to be missing links. Similarly, all observed links are ranked and the links with lower scores are more likely to be the spurious links. The whole network reconstruction algorithm is presented in Algorithm 3.

\begin{algorithm}[!htb]
\caption{Network reconstruction algorithm}
\label{alg::conjugateGradient}
\begin{algorithmic}[1]
\INPUT adjacency matrix ${{\bf{X}}}$ of observed network;
\STATE Obtain the optimal representation matrix ${\bf{Z}^{*}}$ of matrix ${{\bf{X}}}$ via Algorithm 1 or Algorithm 2;

\STATE Construct the adjacency matrix $\bf{SM}$ using (17);

\STATE Separate $\bf{SM}$ into the positive component ${\bf{S}}{{\bf{M}}^ + }$ and the negative component ${\bf{S}}{{\bf{M}}^ - }$ according to the entries' sign;

\STATE Remove the existing entries of ${\bf{X}}$ in ${\bf{S}}{{\bf{M}}^ + }$, and the remained entries with the higher scores are more likely to be the missing links; 

\STATE Sort the existing entries of ${\bf{S}}{{\bf{M}}^ - }$ by comparing it with ${\bf{X}}$, and the ones with the lower scores are more likely to be the spurious links. 

\end{algorithmic}
\end{algorithm}

\textbf{Complexity analysis}. The complexity of Algorithm 1 concentrates on Steps 3-5, in which the SVD (Singular Value Decomposition) is required by the SVT operator in Step 3. Let $n$ denote the size of samples, the cost of decomposition is $O({n^3})$, which is costly for reconstructing large-scale networks. Based on the relaxation in Algorithm 2, the computational complexity for network modeling can be greatly reduced, and networks can be reconstructed efficiently with the guarantee of high accuracy of network reconstruction.

\section{\label{sec:level1} Network Reconstructability Regulation}

The core idea of network reconstructability regulation is to control the systemic regularity of networks, and thereby improve the accuracy of network reconstruction. To regulate the regularity of networks, the proposed self-representation network model is straightforwardly used to analyze the intercommunity of local structures and characterize the roles of network links. The working mechanism of network reconstructability regulation is illustrated in Fig. 4.

\begin{figure}[!htb]
\centering
\includegraphics[width=3.4in,height =2.4in]{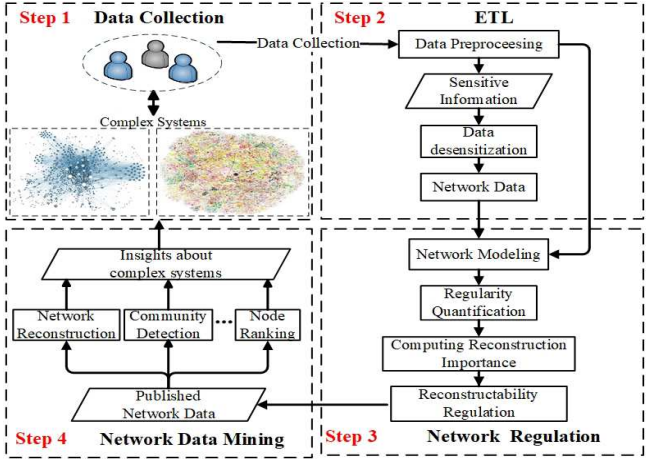}
\caption{Working flowchart of network reconstructability regulation in the context of data publishing. Based on the reconstructability regulation, the privacy, data quality and the performance of downstream tasks can be improved.}
\label{fig_sim}
\end{figure}

\subsection{Network Regularity} \label{sec:model}

Network regularity aims to quantify the extent to which network can be modeled and reconstructed, in other words, can reflect the reconstructability of networks. In regard to regular network, the local structures tend to be represented by the others. Hence, the intercommunity of the local structures can actually reflect the regularity of networks.

According to the discussion in Section IV, the structural intercommunity of local structures contains two aspects. (1) Some local structures of a network may be exactly the same, and the proportion of them can be characterized by the rank of the learned representation matrix $\bf{{{\bf{Z}}^*}}$. The more the number of the identical local structures in the network, the less the rank of the representation matrix. (2) Even though there is no identical local structures, some local structures can also be constructed as the combination of other ones. The more regular the local structures are, the fewer other local structures are needed to represent them, which correspond to the nonzero entries of the representation matrix. So, we apply self-representation network model into networks and measure network regularity based on the representation matrix $\bf{{{\bf{Z}}^*}}$. Based on the aforementioned discussion, we define the network regularity ${\sigma _r}$ as follows.
\begin{spacing}{0.7}
\begin{equation}\label{nt}
{\sigma _r} = \frac{1}{{\sqrt {(n - r)/n} \sqrt {a/(n \cdot r{\rm{)}}} }}
\end{equation}
\end{spacing}
\noindent where $(n - r)/n$ (the fraction of the difference between the dimension $n$ and rank $r$ of $\bf{{{\bf{Z}}^*}}$) denotes the proportion of the identical local structures in the network. The nonzero density $a/(n \cdot r{\rm{)}}$ of the reduced echelon form of matrix $\bf{{{\bf{Z}}^*}}$ (by using Gauss elimination) characterizes the regularity of local structures, where $a$ is the number of the nonzero entries. 
For example, in Fig. 3, the dimension $n$ of matrix $\bf{{{\bf{Z}}^*}}$ equals to 11, and node 3, 10, 11 have identical local structures. Thus, the rank $r$ of $\bf{{{\bf{Z}}^*}}$ must be less than 9. Moreover, we find from Fig. 2 that the local structure of node 1 is more regular than that of node 2, and its reconstruction complexity is lower than that of node 2.
%


\subsection{Reconstruction Importance} \label{sec:model}

By applying the self-representation network model in a network, the learned representation matrix ${\rm{\bf{Z}}}^*$ can reflect the influence of nodes in network reconstruction. Specifically, the $k^{th}$ row of matrix ${\rm{\bf{Z}}}^*$ represents the contribution of node $k$'s local structure for the reconstruction of the other ones. The greater the number of the nonzero entries in the $k^{th}$ row, the higher the frequency of node $k$'s local structure being used for reconstruction. So, we can use the rows of matrix ${\rm{\bf{Z}}}^*$ to estimate the reconstruction importance of network nodes. For node $k$, its reconstruction importance can be defined as follow:

\begin{spacing}{0.7}
\begin{equation}\label{nt}
RC(k) = \frac{1}{n}  \sum\limits_{i = 1}^n {|{\rm{Z}}_{k,i} |}
\end{equation}
\end{spacing}

\noindent where $n$ is the dimension of the $k^{th}$ row. Taking Fig. 2 as an example, according to the representation matrix ${{\bf{Z}}^{\rm{*}}}$, we can conclude that node 5 participates in the reconstruction of four nodes and is most important, and node 1, 3, 7, 9 participate in the reconstruction of only two nodes and are less important. 

\begin{figure}[!t]
\centering
\includegraphics[width=3.0in,height =1.2in]{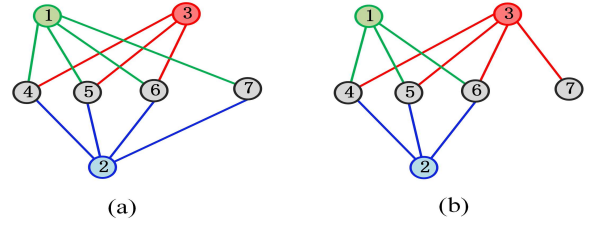}
\caption{The intercommunity between local structures. (a) Missing links prediction; (b) Spurious links identification.}
\label{fig_sim}
\end{figure}

According to the self-representation network model, network regularity is actually reflected by the intercommunity of local structures. The low-rank pursuit and error minimization constraint collectively require the proposed network model to reconstruct node's neighborhood to increase the number of similar local structures while add noisy data as few as possible.
Taking Fig. 5 for example, by comparing node 3's local structure with that of node 1 and 2, we can find that adding link $(3,7)$ is more beneficial than deleting links $(1,7)$, $(2,7)$ in Fig. 5(a) and deleting link $(3,7)$ is more beneficial than adding links $(1,7)$, $(2,7)$ in Fig. 5(b). According to the structural consistency across the local structures, link $(3,7)$ is likely to be a missing link in Fig. 5(a) and a spurious link in Fig. 5(b). If we perturb the network based on it, i.e., adding link $(3,7)$ in Fig. 5 (a) and removing link $(3,7)$ in Fig. 5(b), the network will become more regular. Therefore, the network regularity can be regulated by network links based structural perturbation. So, how to identify the roles of network links and measure their importance in terms of network regularity becomes an important problem.


\begin{algorithm}[!htb]
\caption{Structure perturbation based network reconstructability regulation}
\label{alg::conjugateGradient}
\begin{algorithmic}[1]
\INPUT adjacency matrix ${{\bf{X}}}$ of network $G$;
\OUTPUT  the regulated network $G^R$;

\STATE Obtain the optimal representation matrix ${\bf{Z}^{*}}$ of the observed network using Algorithm 1;

    \FOR{each network node $k{\in}{V}$} 
        \STATE Calculate the importance $RC(k)$ of node $k$ using (19);
    \ENDFOR

    \FOR{each network link $(i,j){\in}{E}$}
        \STATE Calculate the importance $U_{ij}$ of link $(i,j)$ using (20);
    \ENDFOR

\STATE Sort the importance scores $\{U_{ij}\}$ in ascending order and get ranked list $U$ in which the links with the lower scores are more likely to be irregular links;
\STATE $i=0$;
\WHILE {network regularity is increased}
\STATE Removing the irregular network link $U[i]$;
\STATE $i=i+1$;
\STATE Calculate the regularity ${\sigma _r}$ of resulted network using Formula (18);
\ENDWHILE
\STATE \textbf{output} the resulted network.
\end{algorithmic}
\end{algorithm}

By analyzing the learned representation matrix $\bf{{{\bf{Z}}^*}}$, we can find that there are some links that participate frequently in the network self-representation and others that are rarely used. Thus, network links may have different functions in network reconstruction. According to the reconstruction importance of network nodes defined in Formula 19, the greater value of a node's reconstruction importance means the higher frequency of the node's related links being used for network reconstruction. Then, the importance of network links can be estimated by the reconstruction importance of its end nodes, i.e.,
\begin{equation}\label{nt}
{U_{ij}} = RC(i) \times RC(j)  
\end{equation}
The metric quantifies the potential influence of link $(i,j)$ in both directions. The links with the greater value of ${U_{ij}}$ are more likely to be the regular links. Otherwise, they are more likely to be the irregular links. By employing the reconstruction importance of network links, network reconstructability can be regulated by important links based structure perturbation. For example, network reconstructability can be improved by removing the irregular links with low values of $U_{ij}$. The whole algorithm of structure perturbation based network reconstructability regulation is given in Algorithm 4.



\section{Experiments}

We conduct experimental study of the proposed algorithm based on real-world networks. Three sets of experiments are performed to evaluate the following performance:

(1) the effectiveness and efficiency of the proposed network reconstruction algorithm;

(2) the effectiveness of the irregular links identification based network reconstructability regulation;

(3) the effectiveness of the network regularity metric.

\subsection{Experimental Setup}

We consider the following 10 real-world networks drawn from disparate fields: (i) Jazz~\cite{gleiser2003community}, a collaboration network of jazz musicians; (ii) Worldtrade~\cite{Smith1992Structure}, the network of miscellaneous manufactures of metal among 80 countries in 1994; (iii) Contact~\cite{kunegis2013konect}, a contact network between people measured by carried wireless devices; (iv) Metabolic~\cite{duch2005community}, a metabolic network of C.elegans; (v) Mangwet~\cite{baird1998assessment}, the food web in Mangrove Estuary during the wet season; (vi) Macaque~\cite{da2007predicting}, the cortical networks of the macaque monkey; (vii) USAir~\cite{Batageli}, the US Air transportation network; (viii) Facebook~\cite{viswanath2009evolution}, a directed network of a small subset of posts to other user's wall on Facebook. Here we treat it as simple graph by ignoring the directions and weights; (ix) Router~\cite{spring2002measuring}, a symmetrized snapshot of the structure of the Internet at the level of autonomous systems; (x) Yeast~\cite{bu2003topological}, a protein-protein interaction network in budding yeast.

To evaluate the performance of the reconstruction algorithms, we adopt two standard metrics, i.e., AUC (Area Under the Receiver operating characteristic curve) and Accuracy. Among $n$ times of independent comparisons, if there are $n'$ times in which the score of the missing (spurious) link is higher (or less) than that of the non-existent (existent) link and $n''$ times in which the two have the same score, then AUC can be calculated by ${\bf{AUC}}{\rm{ }} = {\rm{ }}{\bf{n}}' + {\bf{0}}.{\bf{5n}}''/{\bf{n}}$. If all the scores are generated from an independent and identical distribution, AUC will approximate to 0.5. Therefore, the extent to which AUC exceeds 0.5 indicates how much better the algorithm performs than the pure chance. Accuracy is defined as the ratio of the relevant links to the number of the selected links. If $L_p$ links among the top-$L$ links are accurately predicted, then ${\bf{Accuracy}}{\rm{ }} = {\rm{ }}{\bf{L_p}}{\rm{ }}/{\rm{ }}{\bf{L}}$.

For comparison, we introduce six benchmark methods. The simplest is the common neighbor (CN)~\cite{liben2007link} in which two nodes have a higher connecting probability if they have more common neighbors. An improved method based on CN is the resource allocation (RA)~\cite{zhou2009predicting}, which assigns less-connected neighbors more weight. Unlike the above two local methods, the LP method~\cite{zhou2009predicting,Lv2009Effective} utilizes quasi-local topological information by summing over the collection of paths with length 2 and 3. In addition, we compare our reconstruction algorithms with three global methods, including SPM~\cite{L2015Toward}, NMF~\cite{Menon2011Link}, and RPCA~\cite{pech2017link}. For convenience, we denote the low rank representation (Algorithm 1) based reconstruction algorithm as LRNR and the low Frobenius-norm representation (Algorithm 2) based reconstruction algorithm as LFNR.

\begin{table*}[!htb]\small \centering
\caption{AUC (top half) and Accuracy (bottom half) of the network reconstruction algorithms for missing links prediction. Each value is averaged over 20 independent runs with 10\% random links as probe set. The parameters of the methods are tuned to their optimal values. The best results are emphasized in bold, and the values in bracket are the standard deviation. }
\includegraphics[width=7.0in]{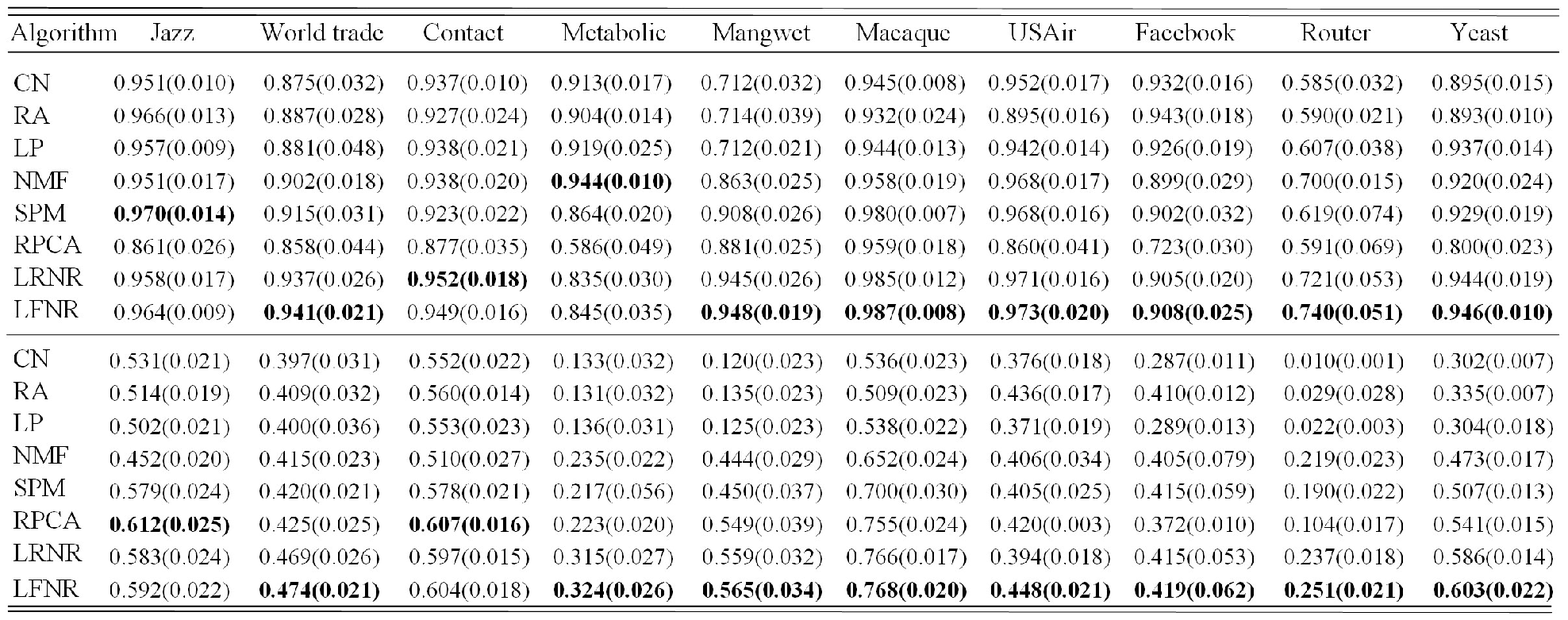}
\end{table*}

\begin{table*}[!htb]\small \centering
\caption{AUC (top half) and Accuracy (bottom half) of the network reconstruction algorithms for spurious links identification. Each value is averaged over 20 independent runs. The parameters in NMF, RPCA, LRR and LFR are tuned to their optimal values. The best results are emphasized in bold. The values in bracket are the standard deviation. }
\includegraphics[width=7.0in]{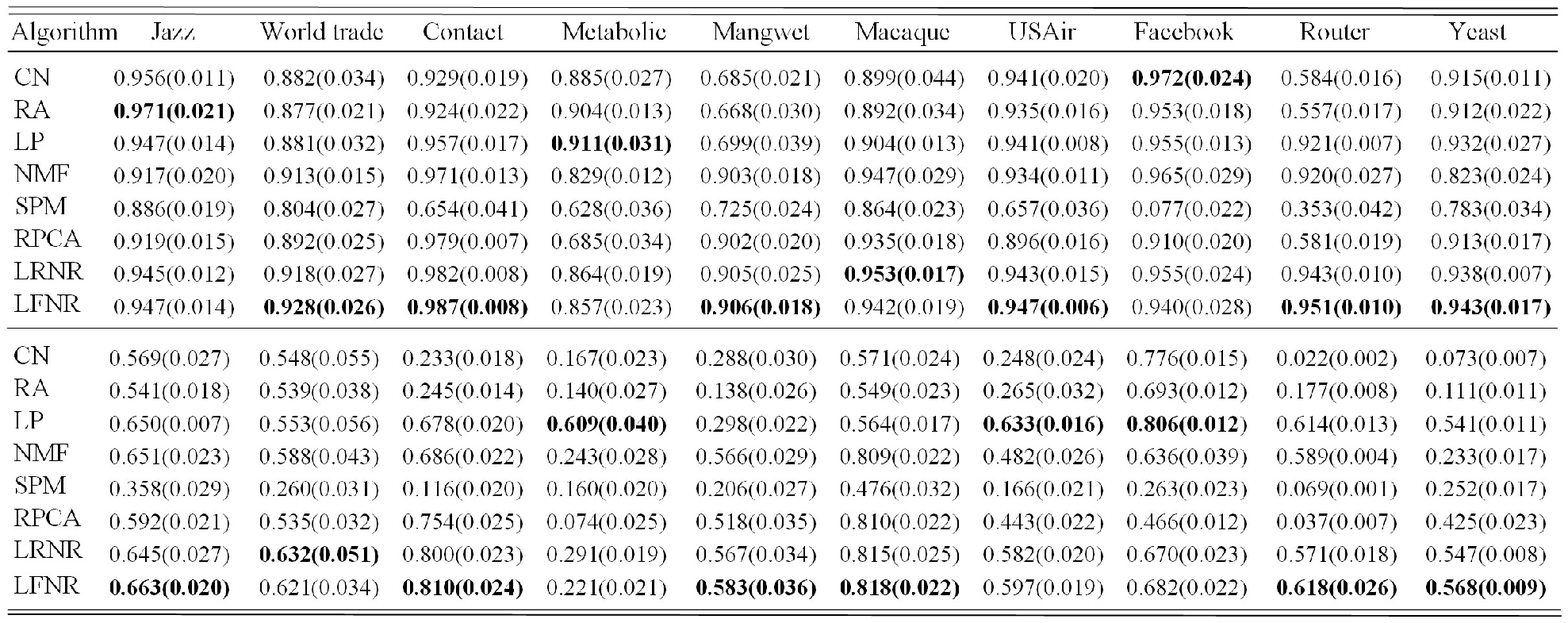}
\end{table*}

\begin{table*}[!htb]\small \centering
\caption{ Time performance of the proposed network reconstruction algorithms (unit: second).}
\includegraphics[width=7.0in]{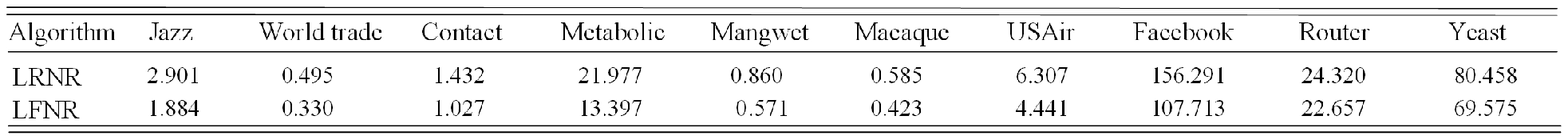}
\end{table*}

\begin{figure*}[!htb]\small \centering
\centering \subfigure[Identified irregular links for network reconstructability regulation.]
{ \includegraphics[width=6.8in,height =1.7in]{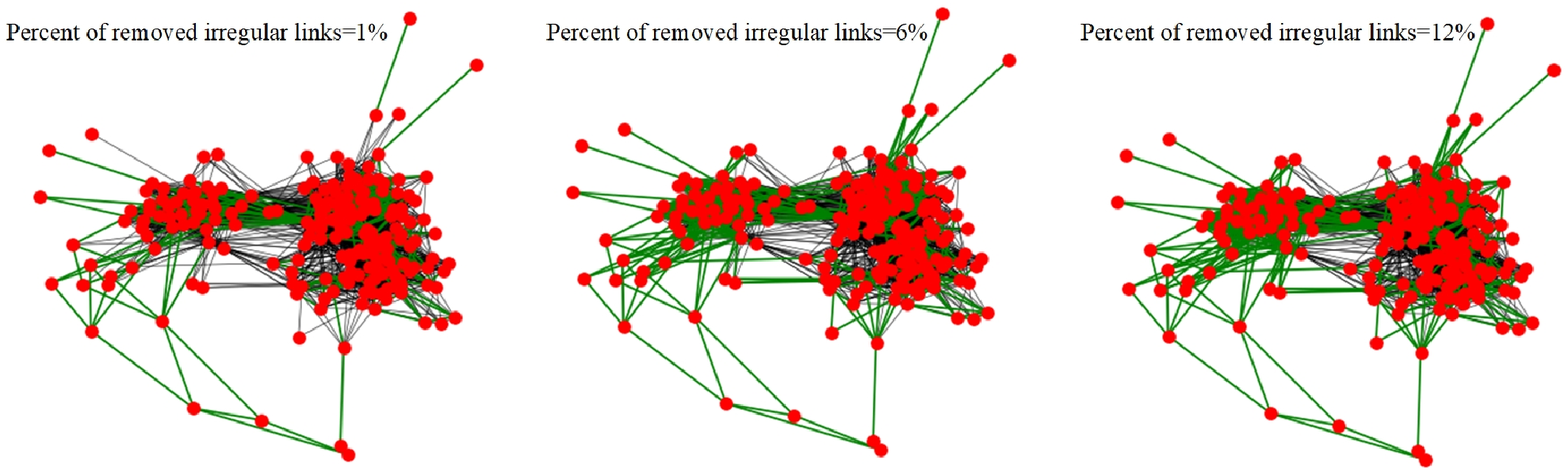}}
\centering  \subfigure[Selected random links for network reconstructability regulation.]
{\includegraphics[width=6.8in,height =1.7in]{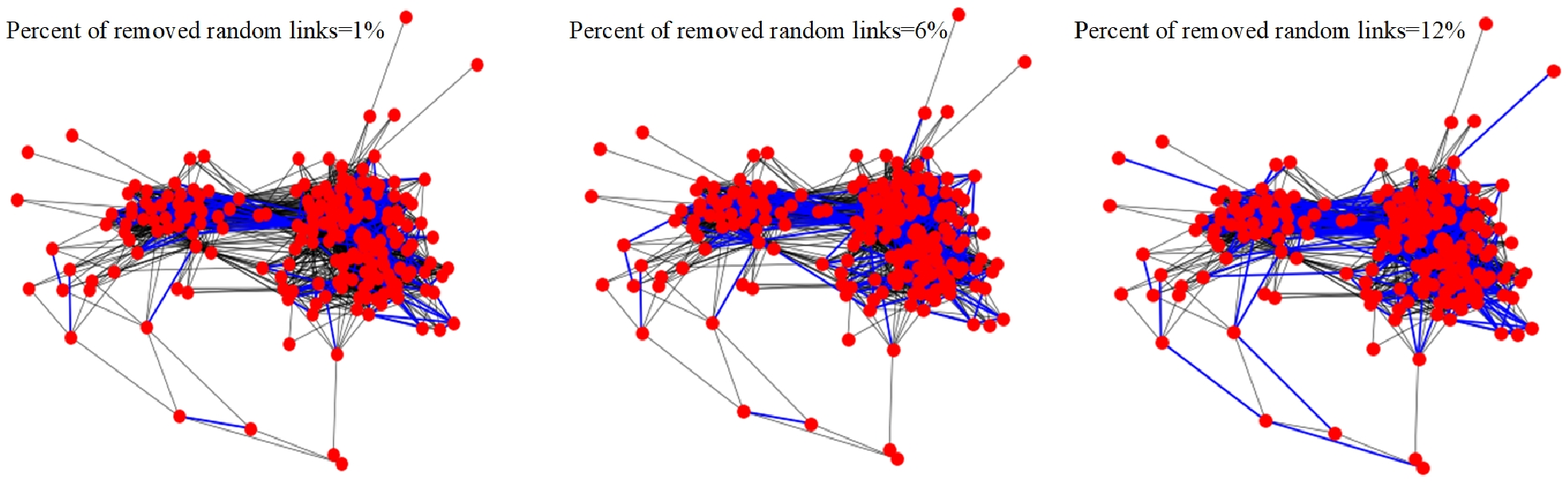}}
\centering  \subfigure[ Reconstructability regulation of Jazz evaluated with different reconstruction algorithms.]
{\includegraphics[width=7.0in,height =1.8in]{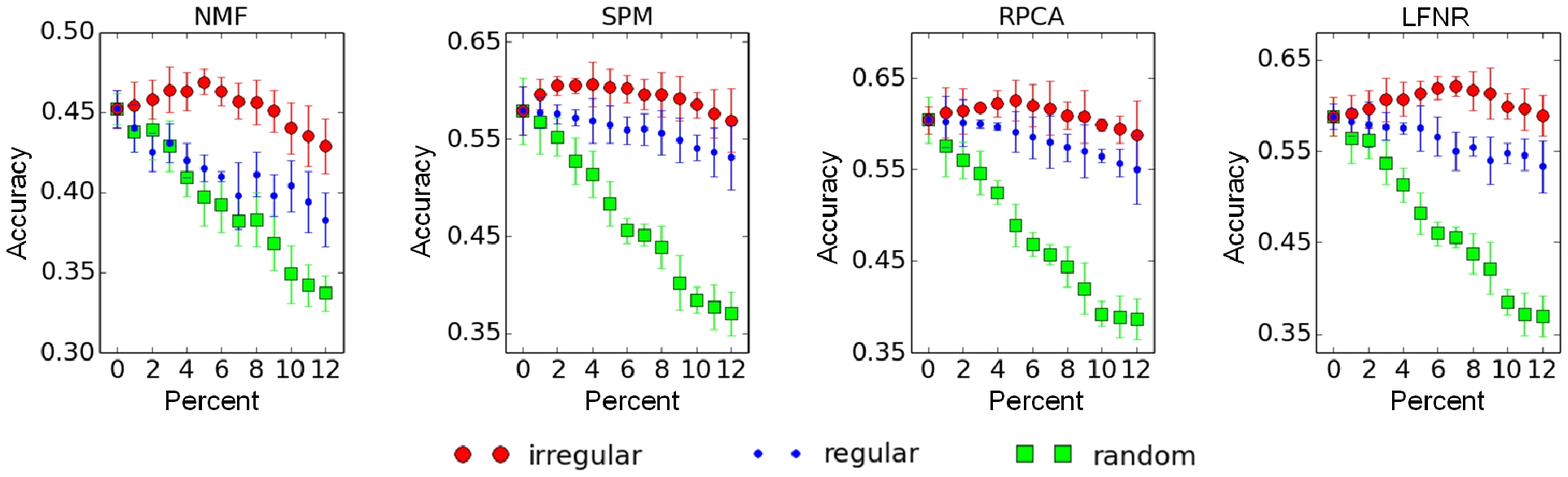}}
\caption{Reconstructability regulation of network Jazz. The links with green color in (a) are the irregular links selected based on the proposed reconstruction importance. The links with blue color in (b) are the links selected randomly. The result in (c) shows the accuracy of network reconstruction algorithms with the removed network links increased from 1\% to 12\%.
}
\end{figure*}

\subsection{Network Reconstruction Evaluation}

To test the validity of the network reconstruction algorithms, we select $10\%$ of the network links as the missing link set $E^M$ (the probe set) and use the remaining $90\%$ as training set $E^T$. The results of missing links prediction measured by AUC and the Accuracy are shown in Table 2. All the data points are obtained by averaging over 20 implementations with independently random division of training set and missing link set. For every network, the bold number in the corresponding column emphasizes the highest accuracy. According to the results in Table 2, we can conclude that the proposed LFNR method generally performs the best among the state-of-the-art algorithms and LRNR performs slightly worse for missing links inferring.

To evaluate the effectiveness of network reconstruction algorithms for spurious links identification, $10\%$ spurious links (the probe set) are added randomly into every real network to construct observed network. The results for spurious links identification measured by AUC and Accuracy are shown in Table 3. For all the networks, our method LFNR performs the best among the state-of-the-art algorithms, usually remarkably better than the second best. The possible reason is that the low-rank and sparse model adopted by this study has a greater expressive capability than the other methods.

To further verify the advantage of LFNR over LRNR for network reconstruction, we also analyze the time performance of them and the results are shown in Table 4. We can find from Table 4 that the LFNR algorithm offers a significant performance improvement in running time and achieves at least $30\%$ complexity reduction.

\subsection{Network Reconstructability Regulation Evaluation}

\begin{figure*}[!htb]
\centering
\includegraphics[width=7.0in,height =3.0in]{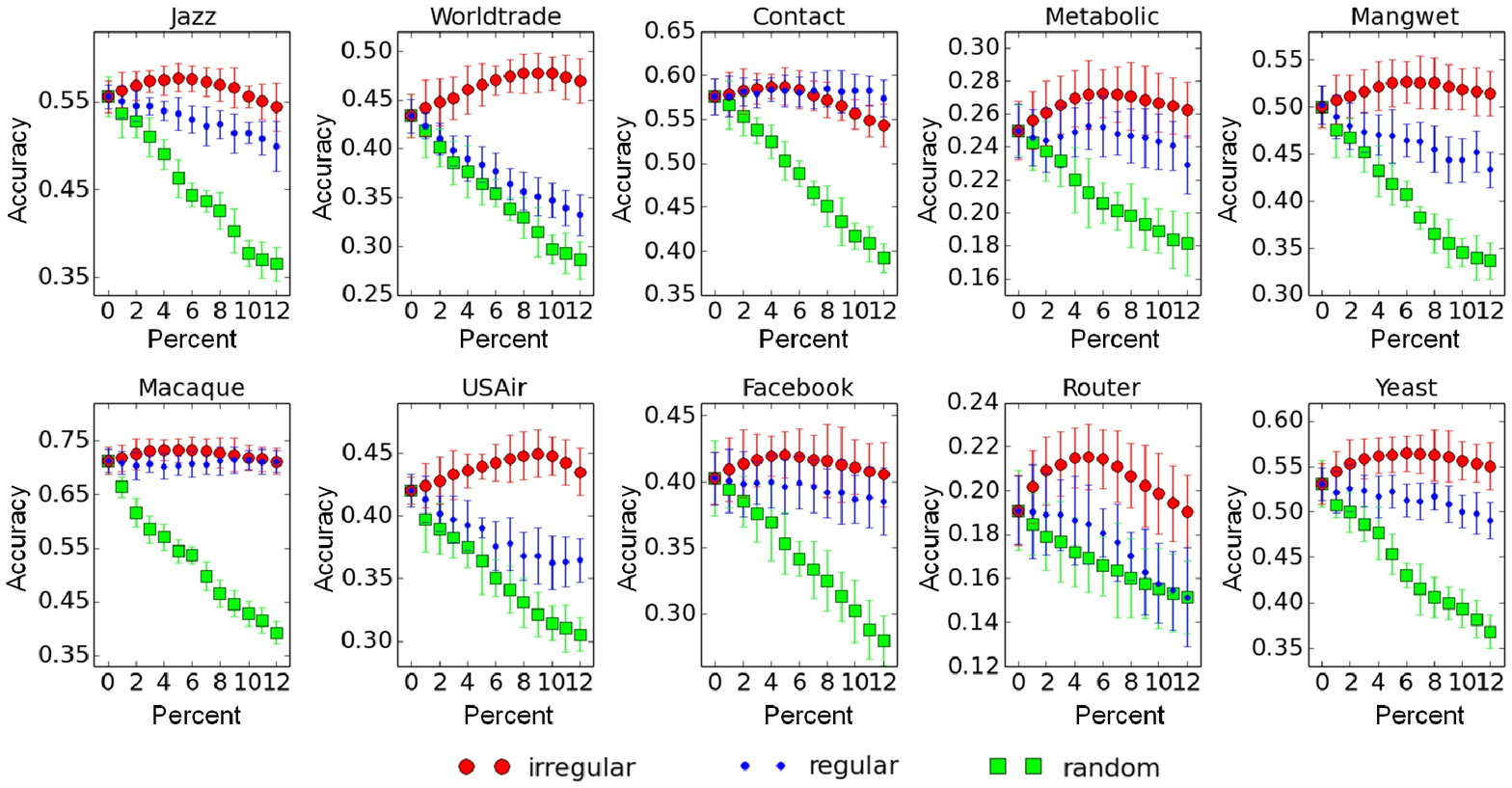}
\caption{Reconstructability regulation (from the perspective of average accuracy) with network links based structure perturbation. The results show the accuracy of the network reconstruction algorithms with the removed network links increased from 1\% to 12\%. Each value is the average over the reconstruction methods NMF, SPM, LRNR and LFNR. The error bars represent the standard deviations of the reconstruction accuracy.}
\label{fig_sim}
\end{figure*}

To evaluate the effectiveness of the proposed network reconstructability regulation algorithm, we firstly identify irregular network links based on the proposed reconstruction importance metric and then remove them to improve network's regularity level. The obtained network is expected to have a higher reconstructability than the original network, which can be quantified by the accuracy of reconstruction, irrespective of the specific network reconstruction algorithm. To validate the rationality of the defined reconstruction importance metric for target links selection, here we adopt two other strategies, i.e., regular link selection and random link selection, as baselines. Specifically, the observed network links are ranked based on reconstruction importance in descending order and the top ones are selected as regular links, and a set of network links are selected randomly as random links.

In order to analyze the performance of the structure perturbation based network reconstructability regulation algorithm in detail, we apply the algorithm in the Jazz network and the results are shown in Fig. 6. Fig. 6(a) presents the identified irregular network links (in green color) of percent 1, 6 and 12, respectively, and Fig. 6(b) are the selected random links (in blue color) of percent 1, 6 and 12, respectively. Compared with the links selected randomly in Fig. 6(b), the identified irregular links in Fig. 6(a) are more likely to be the weak links between the periphery nodes of the network. This difference is obvious between the first network of Fig. 6(a) and that of Fig. 6(b). One explanation about the preference in irregular links identification is that the excessive sparsity of the local structures of the periphery nodes makes they cannot form regular structural patterns and tend to be categorized as irregular elements. With the increase of the degree of network reconstructability regulation, more periphery links are selected as irregular links. To the regulated networks with varied perturbation ratio, the reconstruction results based on NMF, SPM, LRNR and LFNR are shown in Fig. 6(c). We can find from Fig. 6(c) that there is a range in which the reconstruction accuracy can be improved via the removing of irregular links.


To verify the effectiveness of the proposed network reconstructability regulation algorithm, all the real networks are regulated with varied perturbation ratio and result in regulated networks with various regularity. In every regulated network, NMF, SPM, LRNR and LFNR are used for network reconstruction, and the average reconstruction accuracy of the algorithms under various percent of removed links are shown in Fig. 7. As shown in Fig. 7, the accuracy of reconstruction can be improved by removing of irregular network links. The improvement implies that the the structural regularity of networks can be strengthened by irregular links based structure perturbation, and it is an effective way to optimize the reconstructability of networks. As the number of removed links continues to grow, the sparsity of the networks would increases, which would have an adverse influence on the accuracy of network reconstruction.
In contrast, by applying the regular link selection and random link selection strategies, the accuracy of reconstruction degrades continuously with the percent of the removed links increases. Moreover, we can find from Fig. 7 that the regular links based perturbation has less adverse influence on the accuracy of reconstruction than the random links based perturbation. The reason behind this is that the regular links generally have more equivalent links than the random links and the removing of the regular links has less impact on the regularity level of networks. 

Real networks often have different levels of regularity, and the existence of irregular links is a common phenomenon. The different proportion of irregular components causes that not all the networks can be improved remarkably via irregular links based perturbation. The results in Fig. 7 show that the obvious enhancement of the accuracy of network reconstruction often occurs in the networks having low reconstruction accuracy, and the reconstructability of the networks producing high accuracy of network reconstruction is hard to be optimized. Lastly, we can conclude that there is no much room for reconstructability improvement in networks with high regularity and the proposed reconstructability regulation algorithm is more suitable for irregular networks.

\subsection{Performance Evaluation of Network Regularity}

\begin{figure*}[!htb]
\centering \subfigure[Network regularity of Jazz network.]{ \label{fig:side:a}
\includegraphics[width=3.1in,height =2.5in]{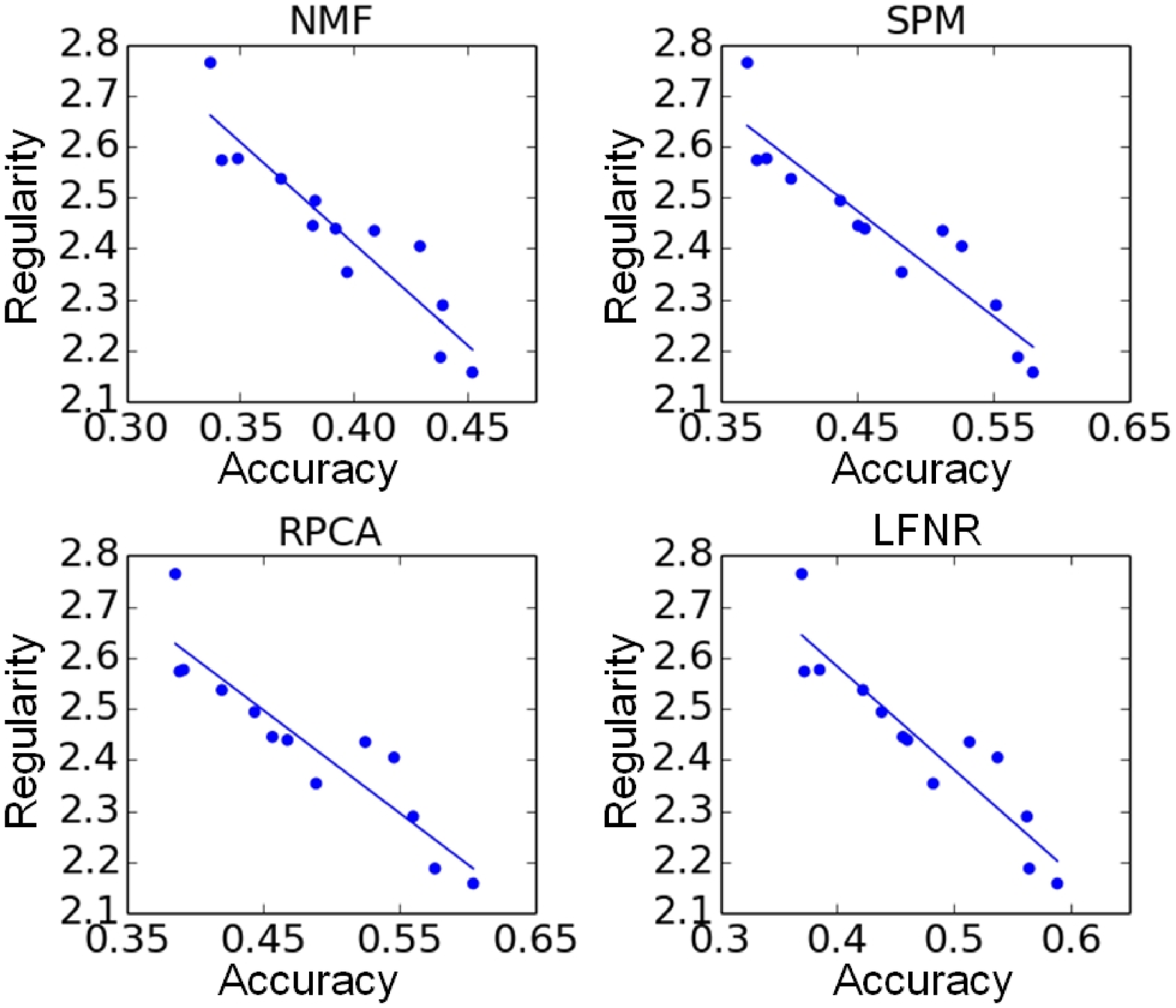}}
\centering  \subfigure[Structural consistency of Jazz network.]{ \label{fig:side:b}
\includegraphics[width=3.1in,height =2.5in]{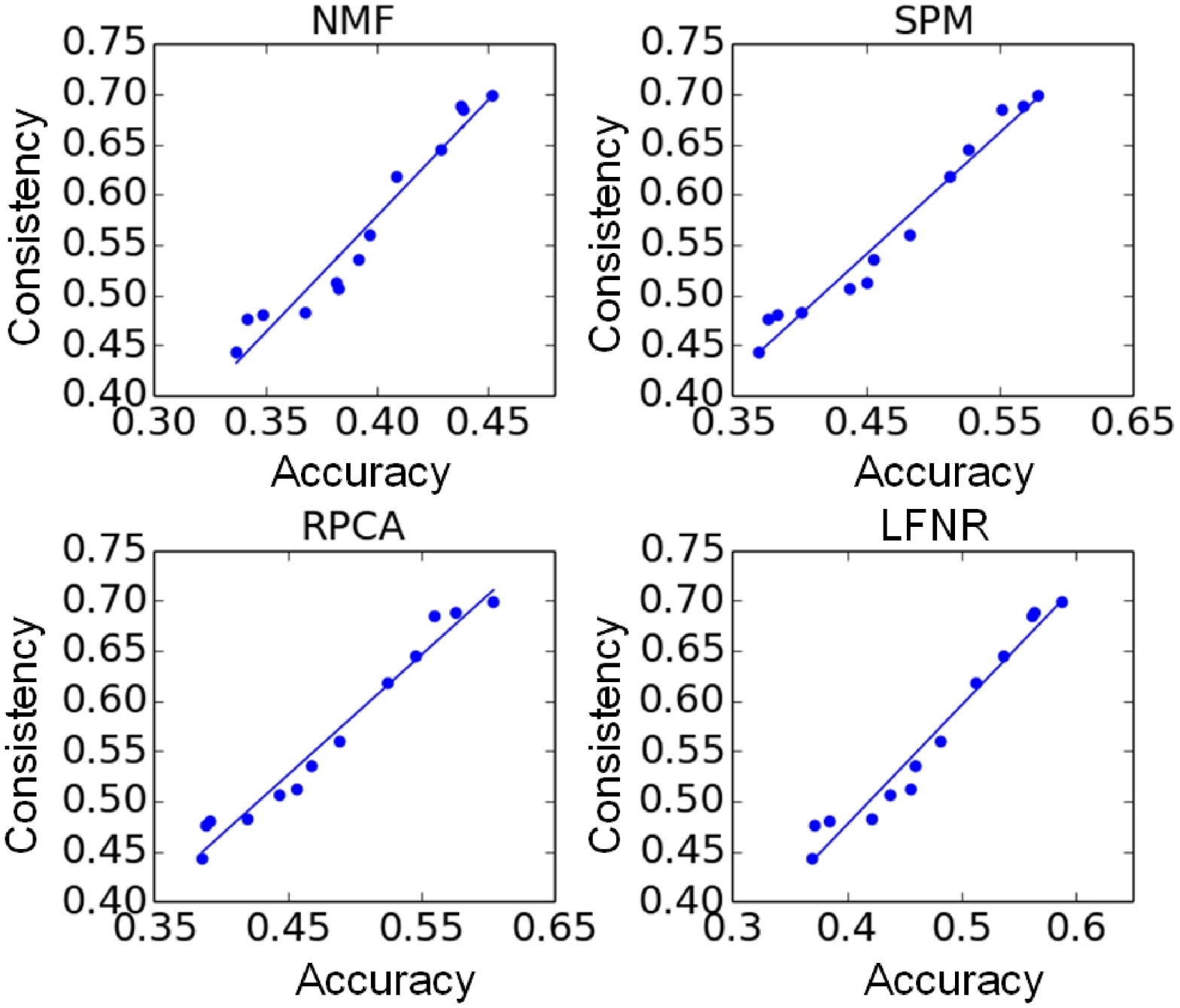}}
\centering  \subfigure[Network regularity of real-world networks.]{ \label{fig:side:b}
\includegraphics[width=6.8in,height =2.6in]{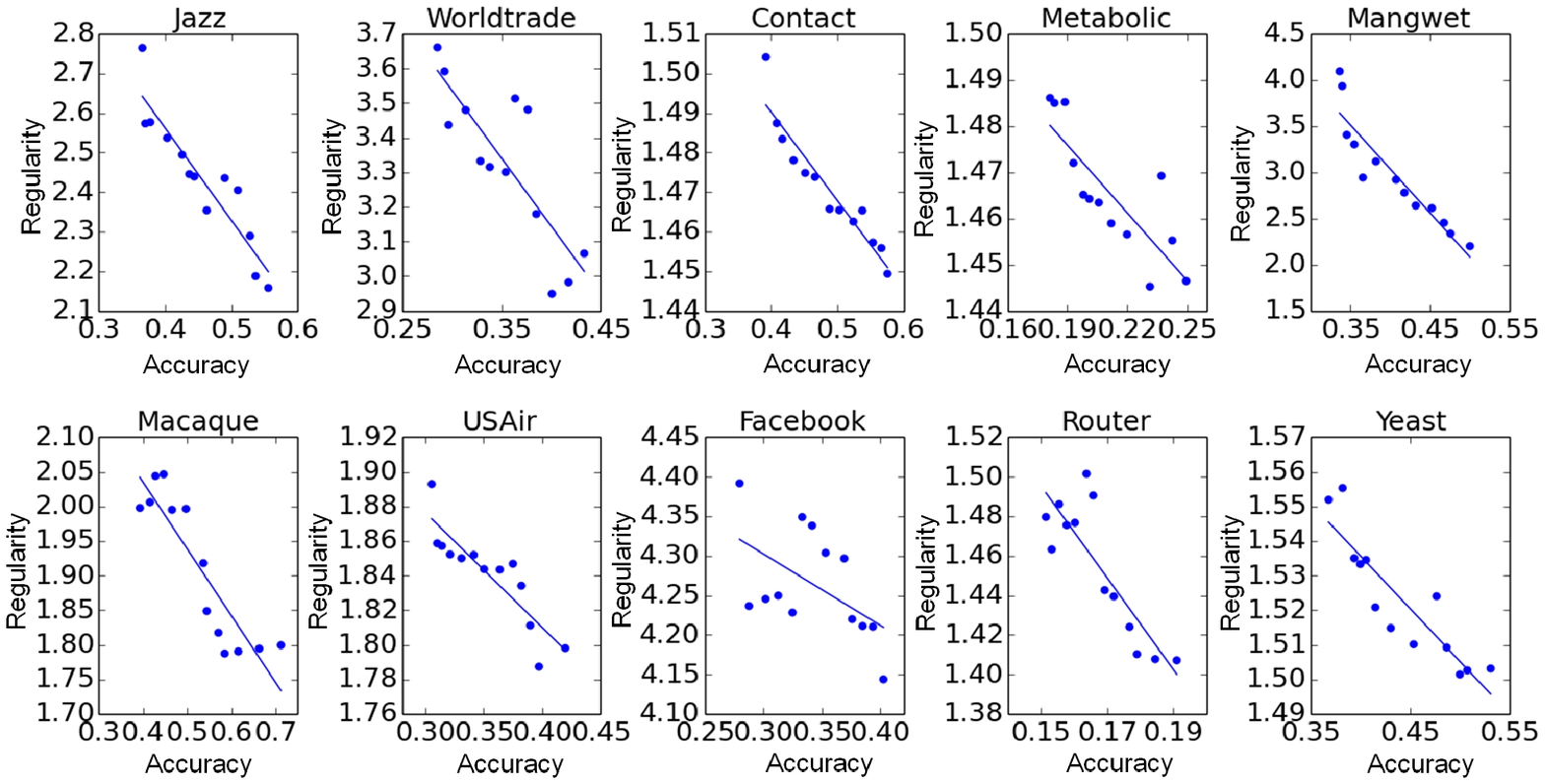}}
\caption{Scatter plot between network regularity and reconstruction accuracy. The x axis is the structural regularity and the y axis is the reconstruction accuracy of networks perturbed with varied percent of removed irregular links. For comparison, the detailed structural regularity and structural consistency of network Jazz are calculated and shown in (a) and (b) respectively. The solid lines are the linear fittings of the results. The smaller the values of structural regularity, the more regular the networks are. Each accuracy value in (c) is the average over the algorithms NMF, SPM, LRNR and LFNR.}
\label{fig_sim}
\end{figure*}

The reason behind the different reconstruction accuracy of the networks is that they possess diversified structural regularity. So, structural regularity becomes a critical property for network analysis. To evaluate the effectiveness of the defined network regularity metric, we compare the values of the regularity metric with the reconstruction accuracy of the four representative network reconstruction algorithms. Firstly, we perturbs the Jazz network by removing different percent of irregular links, from $1\%$ to $12\%$, to generate multiple regulated networks with different levels of regularity. To every regulated network, we estimate its regularity level and calculate its reconstruction accuracy based the four representative algorithms.

The scatter plot between the network regularity values and the reconstruction accuracy of the Jazz network is shown in Fig. 8(a). We can find that the higher the reconstruction accuracy, the smaller the regularity value (a higher level of network regularity). For the purpose of comparison, the structural consistency [12] of the Jazz network under different degrees of perturbation is also presented in Fig. 8(b). We can observe that there is a positive correlation between the structural consistency and the accuracy of network reconstruction. Therefore, the network regularity and the structural consistency all can be used to indicate the reconstructability of networks. However, there is actually a phase of link prediction in the structural consistency calculation while the network regularity is calculated by mining the network structure directly.

In order to evaluate the proposed network regularity metric adequately, we apply it in all the real networks. The experimental results are given in Fig. 8(c), in which each value is the average over the results generated from NMF, SPM, LRNR and LFNR. In general, the value of network regularity is correlated with the average accuracy of reconstruction in the networks, indicating that the higher regularity level will result in the greater reconstructability of networks. Thus, the results on the networks verify the effectiveness of the proposed network regularity metric ${\sigma _r}$, and the metric can be used to characterize the reconstructability of complex networks.

\section{\label{sec:level1} Conclusions and Discussion}

In reality, various networks are regular at different degrees and network structure can be modeled and reconstructed based on the regularity of them. In this study, a low-rank pursuit based self-representation network model is proposed. Based on this model, networks can be represented as the linear combination of local structures. The model allows us to explore the organization principle of networks by analyzing the role of local structures in global network organization. In addition, the principal structural features can be uncovered via the low-rank pursuit of the model. As the errors and incompleteness in data collection make the observed networks noisy and unreliable, it is of great significance to infer the ``true'' underlying networks based on them. Assuming the noise and errors have not significantly change the structural features of the networks, the defined self-representation network model is applied for pattern learning and network reconstruction, consequently, obtaining satisfactory reconstruction accuracy by compared with the state-of-the-art algorithms. A more practical and efficient network reconstruction algorithm is proposed by relaxing the low-rank constraint. In addition to reconstructing the underlying networks, the self-representation network model naturally incorporates a framework for network regularity analysis. We use the self-representation model to identify the irregular links that are inconsistent with the structural pattern across local structures. After discovering the irregular links, the regularity of networks could be enhanced by structure perturbation based on them, thereby, improving the accuracy of network reconstruction. Meanwhile, the regularity level of networks can be calculated based on the learned representation matrix, which is proved to be an effective indicator of network reconstructability.

It is interesting to find that such norm regularization based models have also been applied to drug development~\cite{pech2018} and graph mining~\cite{pech2018link}. Our work is different from the methods, because the current work adopts different matrix norms for network reconstruction and emphasizes the regularity measuring and regulating of networks. By removing weak edges and enhancing real connections, a dynamic diffusion process based network enhancement (NE) method~\cite{wang2018network} is also developed very recently, which can reflects the significance of our work exemplarily. The approach proposed in this study is in a sense preliminary, and it is possible to develop other effective methods for structure-mining-based network organization principle learning and structural regularity regulation.

\section*{Acknowledgment}

This work is partially supported by the National Key Research and Development Program of China under Grant Nos. 2018YFB0904900, 2018YFB0904905; the National Natural Science Foundation of China under Grant No. 61772091, 61772098; the Sichuan Science and Technology Program under Grant No. 2018JY0448; the Innovative Research Team Construction Plan in Universities of Sichuan Province under Grant No. 18TD0027; the National Natural Science Foundation of Guangxi under Grant No. 2017JJD170122y; the Soft Science Foundation of Chengdu under Grant No. 2017-RK00-00053-ZF; the Scientific Research Foundation for Advanced Talents of Chengdu University of Information Technology under Grant Nos. KYTZ201715, KYTZ201750; the Scientific Research Foundation for Young Academic Leaders of Chengdu University of Information Technology under Grant No. J201701.


\ifCLASSOPTIONcaptionsoff
  \newpage
\fi

\bibliographystyle{IEEEtran}
\bibliography{NetRegularity}

\end{document}